\documentclass[reprint,prl,aps,epsfig,longbibliography,superscriptaddress,twocolumn]{revtex4-2}
\usepackage{graphicx}
\usepackage[unicode=true,colorlinks=true]{hyperref}
\hypersetup{linkcolor=blue,citecolor=blue,urlcolor=blue}
\usepackage{dcolumn}
\usepackage{bm}
\makeatletter
\usepackage{color, colortbl}
\usepackage{mathtools, nccmath}
\usepackage{soul}
\definecolor{Gray}{gray}{0.9}

\newcommand{\Rmnum}[1]{\expandafter\@slowromancap\romannumeral #1@}

\begin{document}
\title{Edge reconstruction of compressible Quantum Hall fluid in the filling fraction range 1/3 to 2/3}

\begin{abstract}
Edge reconstruction of gate-tunable compressible quantum Hall fluids in the filling fraction range 1/3 to 2/3 is studied by measuring transmitted conductance of two individually excited fractional $e^2/3h$ edge modes of bulk 2/3 fractional quantum Hall fluid. Our findings reveal that the measured transmitted conductance deviates from the fully equilibrated value for the filling fraction range 1/3 to 2/3 of the gate-tunable compressible quantum Hall fluids at higher magnetic fields. This observation suggests that at the boundary of the compressible fluid a reconstructed $e^2/3h$ fractional edge mode is present and the mode does not completely equilibrate with the inner dissipative bulk region. Consequently, this outer reconstructed edge mode supports adiabatic charge transport, allowing non-equilibrated current transport through the compressible region. These studies open new avenues for achieving robust fractional edge modes even in compressible quantum Hall fluids under strong magnetic fields, enhancing our understanding of edge state dynamics in these complex systems.
\end{abstract}

\author{Suvankar Purkait} 
\email {suvankar.purkait@saha.ac.in ; \\ suvankarpurkait94@gmail.com}
\affiliation{Saha Institute of Nuclear Physics, 1/AF Bidhannagar, Kolkata 700 064, India}
\affiliation{Homi Bhabha National Institute, Anushaktinagar, Mumbai 400094, India}
\author{Tanmay Maiti}
\affiliation{Saha Institute of Nuclear Physics, 1/AF Bidhannagar, Kolkata 700 064, India}
\affiliation{Homi Bhabha National Institute, Anushaktinagar, Mumbai 400094, India}
\affiliation{Department of Physics and Astronomy,Purdue University,West Lafayette,Indiana 47907,USA}
\thanks{Present affiliation}
\author{Pooja Agarwal}
\affiliation{Saha Institute of Nuclear Physics, 1/AF Bidhannagar, Kolkata 700 064, India}
\affiliation{Homi Bhabha National Institute, Anushaktinagar, Mumbai 400094, India}
\affiliation{SPEC, CEA, CNRS, Université Paris-Saclay, CEA Saclay, 91191 Gif sur Yvette Cedex, France.}
\thanks{Present affiliation}
\author{Suparna Sahoo}
\affiliation{Saha Institute of Nuclear Physics, 1/AF Bidhannagar, Kolkata 700 064, India}
\affiliation{Homi Bhabha National Institute, Anushaktinagar, Mumbai 400094, India}
\author{Sreejith G J}
\affiliation{Indian Institute of Science Education and Research, Pune 411008, India}
\author{Sourin Das}
\affiliation{Department of Physical Sciences, IISER Kolkata, Mohanpur, West Bengal 741246, India}
\author{Giorgio Biasiol}
\affiliation{CNR-IOM – Istituto Officina dei Materiali, 34149 Trieste, Italy}
\author{Lucia Sorba}
\affiliation{NEST, Istituto Nanoscienze-CNR and Scuola Normale Superiore, Piazza San Silvestro 12, I-56127 Pisa, Italy}
\author{Biswajit Karmakar}
\affiliation{Saha Institute of Nuclear Physics, 1/AF Bidhannagar, Kolkata 700 064, India}
\affiliation{Homi Bhabha National Institute, Anushaktinagar, Mumbai 400094, India}

 \maketitle
\maketitle

\section*{Introduction}

Quantum Hall (QH) effect is characterized by quantized Hall conductance as $\nu e^2/h$ \cite{Tsui1982,Prange1987}, where $\nu$ is the Landau level filling fraction. At the well defined QH states, the bulk is insulating and topologically protected edge states of total conductance $\nu e^2/h$ at the boundary carry the current \cite{Halperin1982, ButtikerPRL, ButtikerPRB,chang1990}. From the bulk to the physical edge of the sample, the carrier density reduces to zero because of the surface barrier potential. At the smooth boundary of the two dimensional electron system (2DES), formation of dominant incompressible strips results in integer and fractional edge modes \cite{Beenakker1990,Chklovskii1992,Wen1994,Wan2003}. Depending upon the smoothness of the surface potential and the strength of Coulomb interaction, multiple upstream and downstream fractional charge modes as well as charge neutral modes appear \cite{MacDonald1990,Bid2010,Inoue2014}.

The studies of edge reconstruction and the robust reconstructed edge modes are very crucial for quantum interferometric applications of the edge modes. The impact of equilibration of reconstructed edge modes in the quantum interferometric applications is studied \cite{Bhattacharyya2019,Biswas2024}. The study of equilibration of the edge modes can reveal many useful information, like the interaction between edge modes hosting different quasi-particles \cite{Lin2019,Maiti2021}, the energy gap corresponding to the incompressible strips separating edge modes \cite{Maiti2021,Deviatov2002,Kononov2012}, the estimation of edge tunneling exponent in quantum point contact \cite{Cohen2023,Hu2011,Wan2005,Varjas2013,Wen1990,Joglekar2003,Wan2002} etc.

\begin{figure}[htbp]
\includegraphics[width=8.6 cm, height=5 cm]{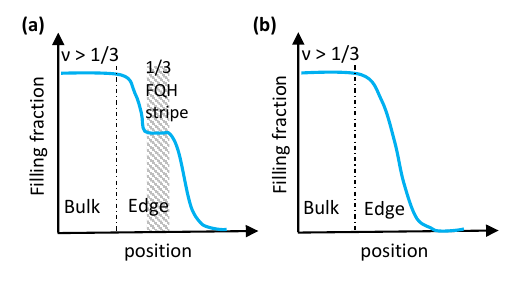}
\centering \caption[ ] 
{\label{fig:schematics of edge reconstruction} (color online) (a) Schematic filling fraction profile at the smooth boundary under electrostatic edge reconstruction for bulk filling fraction $1/3 < \nu < 2/3 $. The dominant incompressible stripe (shaded 1/3 FQH state) separates the outer 1/3 mode from the bulk compressible fluid. (b) Schematic filling fraction profile in absence of electrostatic edge-reconstruction at the boundary. No edge mode is expected in this case.}
\end{figure}

There have been many theoretical \cite{Joglekar2003,Wan2002,meir1994,Ihnatsenka2013,Amaricci2017,Yang2003,Wang2013, Zhang2014,Khanna2021} and experimental \cite{Kouwenhoven1990, Ronen2018,Kumar2022,sabo2017,Nakamura2023prl} studies on edge reconstruction at different incompressible filling fractions $\nu$. Reconstructed fractional edge modes are observed at integer filling unity \cite{Kouwenhoven1990,Bhattacharyya2019,Maiti2020}. Edge reconstruction for 2/3 fractional quantum Hall (FQH) state is well studied. One of the theoretical models for 2/3 FQH state assumes one downstrean $e^2/h$ integer mode and one upstream $e^2/3h$ fractional mode \cite{MacDonald1990} which upon equilibration produce only one downstream $2e^2/3h$ charge mode along with a neutral upstream mode \cite{KFP1994,KF1995}. Other model assumes multiple upstream and downstream edge modes which after inter channel interactions give rise to two co-propagating downstream $e^2/3h$ modes along with two upstream neutral modes  \cite{meir1994,Wang2013}. Subsequent experiment confirms validity of the second model of edge reconstruction for 2/3 FQH state in case of smooth confining potential at the boundary of the 2DES \cite{sabo2017}, where equilibration of the fractional charge modes is found to be about 2 $\mu$m. In other experiment at higher magnetic fields, more robust co-propagating fractional edge modes of conductance $e^2/3h$ are observed with equilibration length of the order $100 \ {\mu m}$ for $\nu = 2/3$ in the smooth  boundary \cite{Maiti2020}. Such a robust 1/3 fractional edge mode at bulk filling fraction 2/5 is used for QH interferometry and studying braiding statistics of fractional charges recently \cite{Kundu_2023}. 

So far edge reconstruction is studied in the incompressible QH states extensively. However, edge reconstruction in the compressible region in between two FQH states is least studied experimentally. Theoretically edge reconstruction is predicted for incompressible QH fluids as well as compressible QH fluids \cite{Wan2002, Wan2003} in presence of strong Coulomb interaction. In this article, we focus on the edge structure of compressible fluid in the filling fraction range 1/3 to 2/3. One possible edge reconstruction model is schematically shown in Figure \ref{fig:schematics of edge reconstruction}(a) for compressible fluids in the filling fraction range of 1/3 to 2/3, considering the dominant incompressible stripe corresponding to 1/3 FQH state and a robust $e^2/3h$ fractional charge mode might emerge \cite{Beenakker1990,Chklovskii1992}. In case of weaker incompressibility of 1/3 FQH state or the sharp boundary, edge reconstruction should not be observed for the compressible fluid. Figure \ref{fig:schematics of edge reconstruction}(b) schematically represents such a case where edge reconstruction is not favorable and only the dissipative bulk conducts. In this article, we present experimental study on edge reconstruction of the gate defined compressible QH fluid in the filling fraction range $1/3 <\nu < 2/3$ to verify the applicability of the above two models. 

In this study we have utilized the knowledge and experimental technique to measure edge equilibration that is used in our previous work \cite{Maiti2020}. Following our previous work, we have selectively excited the two partially equilibrated $e^2/3h$  modes of 2/3 FQH state and use them to probe the edge reconstruction of a gate defined compressible fluid in the filling fraction range $1/3 <\nu < 2/3$. Transmittance of the excited $e^2/3h$ fractional modes through the top gate defined compressible fluid is measured. We find that the measured transmitted conductance deviates from the full equilibration value at higher magnetic field. Our observation is explained by considering a reconstructed $e^2/3h$ fractional edge mode that does not fully equilibrate with the inner dissipative bulk region of the compressible fluid. Therefore, the outer reconstructed $e^2/3h$ fractional edge mode at the boundary of the compressible fluid supports adiabatic charge transport that validates edge reconstruction of compressible fluid in the filling fraction range 1/3 to 2/3 as shown in Figure \ref{fig:schematics of edge reconstruction}(a).

\begin{figure}[htbp]
\includegraphics[width=8.6 cm, height= 14.7 cm]{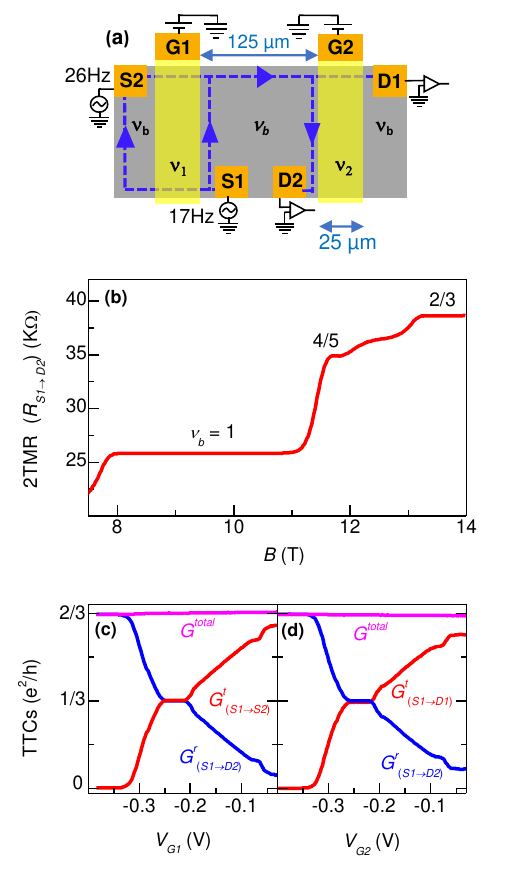}
\centering \caption[ ] 
{\label{fig:device diagram and characterization} (color online) (a) Topologically equivalent device structure with measurement scheme for edge equilibration experiments. S1, S2, D1 and D2 are the Ohmic contacts for current injection and detection; G1 and G2 are the top metal gates (approximate dimensions 25 $\mu$m x 25  $\mu$m) used for tuning filling fractions $\nu_1$ and $\nu_2$ beneath the gates respectively, while bulk is kept at filling $\nu_b$. Sources S1 and S2 are excited simultaneously with different frequencies; currents at virtually grounded D1 and D2 contacts are measured by standard lock-in technique. (b) Plot of two terminal magneto-resistance (2TMR) measured between S1 and D2, when all other contacts are floating and gates are grounded. Quantized plateaus at bulk filling $\nu_b = 2/3, 4/5$ and 1 are labeled. (c) Representative plot of G1 gate characteristics vs $V_{G1}$ at $\nu_b$ = 2/3 ($B =$ 13.7 T) when G2 gate is fully pinched-off. Red curve represents G1 transmittance($G^t_{S1 \rightarrow S2}$) measured between contacts S1 and S2. Blue curve is the reflectance ($G^r_{S1 \rightarrow D2}$) of G1 measured between contacts S1 and D2. Magenta curve represents total conductance ($G^{total}$) i.e. sum of transmittance and reflectance. (d) Plot of G2 gate characteristics at 13.7 T magnetic field, when G1 gate is fully pinched-off. Transmittance $G^t_{S1 \rightarrow D1}$(red curve) is measured between S1 and D1, reflectance $G^r_{S1 \rightarrow D2}$ (blue curve) is measured between S1 and D2. Gate characteristics data represented in (c) and (d) are taken in a separate cooldown for comparison and checking sample uniformity.}  
\end{figure}

\section*{Device and measurement}

A multi-terminal top gated device is made using a modulation-doped GaAs/AlGaAs hetero-structure, where the 2DES resides at the GaAs/AlGaAs heterojunction located 100 nm below the top surface. Figure \ref{fig:device diagram and characterization}(a) represents the schematic device structure, similar to Ref.\cite{Maiti2020}, where Ohmic contacts S1, S2, D1 and D2 are utilized for current injection-detection and two top metal gates G1 and G2 are used to tune the filling fraction underneath the gated region. The device is mounted in a dilution refrigerator equipped with a $14 \ {\rm T}$ superconducting magnet at a base temperature of $7 \ {\rm mK}$. Because of parasitic noise the electron temperature is expected to be higher and in our sample the electron temperature reaches about 30 mK. Carriers in the hetero-structure are injected by light illumination from a GaAs light emitting diode at temperature $3 \ {\rm K}$. The carrier density and low temperature mobility of the sample become $n \sim 2.2\times10^{11} \ {\rm cm^{-2}}$ and $\mu \sim 4\times10^6 \ {\rm cm^2/Vs}$ respectively after light illumination. Transport experiments are done by measuring low frequency ac currents using standard lock-in technique after amplifying the currents with homemade low-noise current to voltage pre-amplifiers. Two-terminal magneto-resistance (2TMR) is measured between contacts S1 and D2 (virtually grounded) with all other contacts open and is shown in Figure \ref{fig:device diagram and characterization}(b). In the two-terminal measurement, robust resistance plateaus at bulk filling $ \nu_b = 2/3, 4/5$ and 1 are observed along with the higher filling QH plateaus (not shown). The $\nu_b = 1$ IQH plateau is observed in the magnetic field range of 8-11 T and $\nu_b=2/3$ plateau starts from B = 13.2 T.

Initially the G1 gate is charterized at bulk filling fraction $\nu_b=2/3$ by varying the gate voltage $V_{G1}$, while G2 gate is kept in fully pinched-off condition. In this characterization, S1 source is excited and the transmitted current is measured at S2 through a customized pre-amplifier as in Ref.\cite{Maiti2020}. The reflected current is measured at D2 contact. The experiments are repeated in two separate cooldowns. A representative G1 gate charactics at B = 13.7 T is shown in Figure \ref{fig:device diagram and characterization}(c) from second cooldown. The sum of transmittance $G^t_{S1 \rightarrow S2}$ and reflecttance $G^r_{S1 \rightarrow D2}$ is the total conductance $G^{total}$ that remains constant showing current conservation. A well developed $e^2/3h$ conductance plateau in the transmittance and reflectance confirms the good quality of the sample beneath the gate. Similarly, G2 gate characteristics is obtained by exciting source S1 and measured transmitted and reflected current at D1 and D2 contacts respectively keeping the G1 gate under fully pinched-off condition. A representative G2 gate characteristics is shown in Figure \ref{fig:device diagram and characterization}(d) for B = 13.7 T. Figure \ref{fig:device diagram and characterization}(c) and (d) shows similar characteristics for both the gates, which confirm the uniformity of the sample.

\begin{figure}[htbp]
\includegraphics[width=8.6 cm, height=7.5 cm]{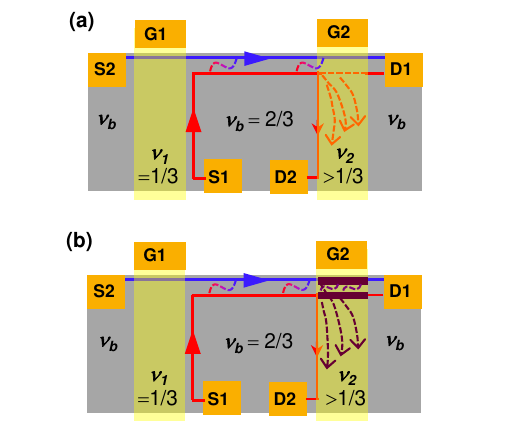}
\centering \caption[ ]
{\label{fig:g1 at 1/3 filling schematics} (color online) (a) Schematic of edge mode connections beneath the gate G2 in the filling fraction range $1/3 < \nu_2 < 2/3$, where a $e^2/3h$ reconstructed edge mode is considered along with a parallel dissipative region. The bulk filling fraction is $\nu_b = 2/3$ and G1 filling fraction is $\nu_1 = 1/3$. Here, blue and red lines indicate the outer and inner 1/3 edge modes respectively. In the schematics only the excited charge modes are shown. The dashed symbols in between two channels indicate equilibration processes. Dashed arrows indicate dissipative current path. (b) Schematic of edge mode connections when reconstructed 1/3 edge mode is not present in the compressible fluid beneath the gate G2. In this case, the currents from S1 and S2 fully equilibrate  beneath the gate G2.}
\end{figure}

To study the edge reconstruction in the filling fraction range of 1/3 to 2/3, two resolved $e^2/3h$ fractional edge modes are required. Therefore, the experiments are carried out at bulk filling fraction $\nu_b$ = 2/3, where the two $e^2/3h$ fractional modes become partially resolved even after co-propagating over 125 $\mu$m distance. Corresponding equilibration length $l_r$ is found to be 104 $\pm 4 \rm \mu m$ at $B = 13.98$ T \cite{Maiti2020}. For separately contacting and exciting the individual fractional edge modes, we follow the scheme laid in Ref.\cite{Maiti2020}, where the gate voltage $V_{G1}$ is tuned to set incompressible FQH states at filling fraction $\nu_1$ = 1/3 beneath the gate G1. By setting $\nu_1$ = 1/3 and exciting the sources S1 and S2 simultaneously with ac 25.8 $\mu V$ at different frequencies $17 \ {\rm Hz}$ and $26 \ {\rm Hz}$ respectively, individual modes are excited separately. In this study, by tuning the filling fraction $\nu_2$ beneath the gate G2 monotonically, the transmitted and reflected conductance through the G2 gate are measured at D1 and D2 respectively by lock-in technique (setup shown in Figure \ref{fig:device diagram and characterization}(a)). The measured two terminal conductances (TTCs) are denoted as $G^{~\nu_1,\nu_2}_{\rm Si \rightarrow Dj}{\rm (\nu_b , B)}$, where i, j = 1, 2 are the indices of the sources and detector contacts respectively. 

The connection to the edge modes in our experimental measurement setup considering edge reconstruction for filling fraction range 1/3 to 2/3 beneath the gate G2 is shown in Figure \ref{fig:g1 at 1/3 filling schematics}(a). In this schematic, the outer $e^2/3h$ edge mode beneath the gate G2 is separated from the inner dissipative region by an incompressible strip corresponding to 1/3 FQH state, as discussed in Figure \ref{fig:schematics of edge reconstruction}(a). In case of non-equilibration during co-propagation over 125 $\mu$m along the mesa boundary, the outer $e^2/3h$ mode will fully transmit from S2 to D1. Under partial equilibration, the non-equilibrated current reaches to D1 through the G2 gated region and the equilibrated current in the inner mode partially reach to D1 and D2 depending on the value of $\nu_2$. In this picture, partial equilibration values of TTCs will be expected.

If there is no edge reconstruction beneath the gate G2 in the filling fraction range 1/3 to 2/3 as discussed in Figure \ref{fig:schematics of edge reconstruction}(b), the outer 1/3 edge mode fully equilibrates beneath the gate G2 as presented in Figure \ref{fig:g1 at 1/3 filling schematics}(b). In this scenario, the currents from S1 and S2 will be fully equilibrated beneath the gate G2. As a result, full equilibration values of TTCs will be expected. We will verify the above two possibilities of edge reconstruction picture in the filling fraction range 1/3 to 2/3 by experimental measurements.

\begin{figure*}[htbp]
\includegraphics[width=18 cm, height=10 cm]{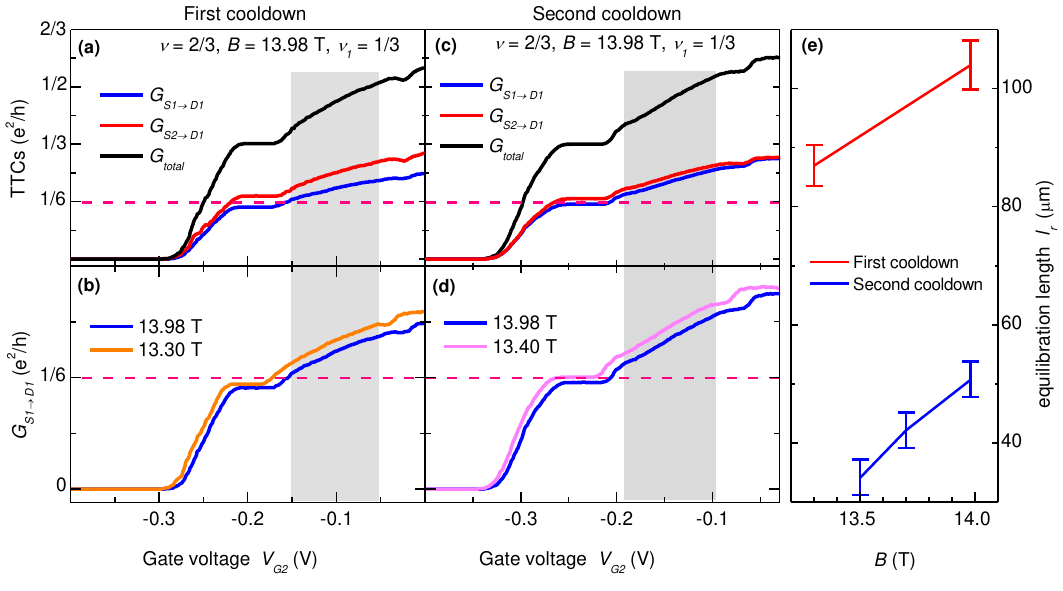}
\centering \caption[ ]
{\label{fig:g1 at 1/3 filling data} (color online) (a) Two terminal conductances (TTCs) are plotted against G2 gate voltage $V_{G2}$ with fixed $\nu_1 = 1/3$ and bulk filling $\nu_b=2/3$ at magnetic field $B$ = 13.98 T for first cooldown. Red curve represents the TTC $G^{~1/3,\nu_2}_{\rm S2 \rightarrow D1}{\rm (2/3, 13.98 T)}$, blue curve describes the $G^{~1/3,\nu_2}_{\rm S1 \rightarrow D1}{\rm (2/3, 13.98 T)}$. Black curve depicts the total conductance, $G_{total}$ measured at D1, that is obtained by adding the blue and red curves. The red dashed horizontal line at $e^2/6h$ value represents the full equilibration value of TTCs at $\nu_2 = 1/3$. The gray shade indicates the compressible region of our interest. (b) Plot of $G^{~1/3,\nu_2}_{\rm S1 \rightarrow D1}{(2/3, B)}$ for magnetic fields $B$ = 13.98 T (blue curve) and 13.30 T (orange curve) of first cooldown. The deviation of $G^{~1/3,1/3}_{\rm S1 \rightarrow D1}$ from full equilibration value (i.e, 1/6 $e^2/h$, red dashed line) is a measure of equilibration length $l_r$ (see text). (c) Plot of TTCs at 13.98 T magnetic field for second cooldown. (d) Comparison of $G^{~1/3,\nu_2}_{\rm S1 \rightarrow D1}{\rm (2/3, B)}$ for magnetic fields 13.98 T (blue curve) and 13.40 T (magenta curve) in second cooldown. (e) Plots of equilibration length $l_r$ with magnetic fields for two cooldowns.}
\end{figure*}
\section*{Results and analysis}
 
The experimentally measured values of TTCs in the first cooldown are presented in Figure \ref{fig:g1 at 1/3 filling data}(a)-(b) with $V_{G2}$ for different magnetic fields. The red curve in Figure \ref{fig:g1 at 1/3 filling data}(a) represents the TTC $G^{~1/3,\nu_2}_{\rm S2 \rightarrow D1}{\rm (2/3, 13.98 T)}$ and the blue curve describes $G^{~1/3,\nu_2}_{\rm S1 \rightarrow D1}{\rm (2/3, 13.98 T)}$. The sum of these two conductances gives the total conductance $G_{total}$, which is plotted in black curve and resembles the G2 gate characteristics. The plot of total conductance $G_{total}$ shows 1/3 conductance plateau, which confirms the formation of robust incompressibility beneath the G2 gated region. The TTCs $G^{~1/3,\nu_2}_{\rm S2 \rightarrow D1}{\rm (2/3, 13.98 T)}$ and $G^{~1/3,\nu_2}_{\rm S1 \rightarrow D1}{\rm (2/3, 13.98 T)}$ do not overlap to each other. The result suggests partial equilibration of the two edge modes during co-propagation. Particularly at $\nu_2 = 1/3$, the two edge modes co-propagate over 125 $\mu$m distance and from the deviation of $G^{~1/3,\nu_2}_{\rm S1 \rightarrow D1}{\rm (2/3, 13.98 T)}$ from the full equilibration value of 1/6$e^2/h$, the equilibration length $l_r = 104 \pm 4 \mu \rm m$ is calculated for magnetic field $B = 13.98$ T using the formula derived in our previous work \cite{Maiti2020}. The interest of the current article is the filling fraction region $1/3 < \nu_2 < 2/3$ as shown by a grey shade in Figure \ref{fig:g1 at 1/3 filling data}, where the TTCs also do not match with each other. The result confirms that the outer mode does not fully equilibrate even after passing through the compressible region below the gate G2. This partial equilibration is consistence with the schematic Figure \ref{fig:g1 at 1/3 filling schematics}(a).

Since the equilibration length increases with increasing magnetic field, further experiments are carried out at lower magnetic field $B = 13.3$ T within the bulk 2/3 QH plateau. In Figure \ref{fig:g1 at 1/3 filling data}(b), the conductances $G^{~1/3,\nu_2}_{\rm S1 \rightarrow D1}{\rm (2/3, B)}$ for $B=$ 13.98 T and 13.30 T are plotted for comparison of equilibration during copropagation. The deviation of conductance from 1/6 at $\nu_2 = 1/3$ becomes lower for $B = 13.3$ T and corresponding equilibration length is estimated as $l_r = 87 \pm 3 \mu \rm m$.

Notably, the total conductance $G_{total}$ in Figure \ref{fig:g1 at 1/3 filling data}(a) does not reach the bulk conductance value of 2/3 at zero gate bias because of little bit lower density below the gate G2. Due to the opacity of the metal top gates, there might be little bit smaller density in the gate defined regions after light illumination for carrier injection. This little density mismatch differs for different cooldown and the 2DES carrier injection conditions. Therefore, the experiments are repeated in a second cooldown of the device. The TTCs data for second cooldown is shown in Figure \ref{fig:g1 at 1/3 filling data}(c) and (d). Measured TTCs for $B=$ 13.98 T is presented in Figure \ref{fig:g1 at 1/3 filling data}(c), where total conductance $G_{total}$ reaches a maximum value about 0.6. However, the difference between $G^{~1/3,\nu_2}_{\rm S2 \rightarrow D1}{\rm (2/3, 13.98 T)}$ and $G^{~1/3,\nu_2}_{\rm S1 \rightarrow D1}{\rm (2/3, 13.98 T)}$ is smaller than the previous cooldown at the same magnetic field value. In the second cooldown, data are collected after several thermal cycle that changed the sample quality. In Figure \ref{fig:g1 at 1/3 filling data}(d) the TTC $G^{~1/3,\nu_2}_{\rm S1 \rightarrow D1}{\rm (2/3, B)}$ for magnetic fields 13.40 T(magenta curve) and 13.98 T (blue curve) are compared. The TTC at 13.40 T magnetic field almost reach the full equilibration value of 1/6$e^2/h$ at $\nu_2 = 1/3$.

The calculated values of the equilibration length at different magnetic fields are plotted in Figure \ref{fig:g1 at 1/3 filling data}(e) for both the cooldowns. In the first cooldown, the equilibration length is seen to be larger compared to the second cooldown. Such lowering of equilibration length in the second cooldown is achieved by thermal cycling of the sample that helps in demonstrating our model at full equilibration. Interestingly, the value of the equilibration length increases with magnetic field for both the cooldown and it is consistent with our previous work \cite{Maiti2020}.

\begin{figure*}[b!htp]
\includegraphics[width=14 cm, height=7.8 cm]{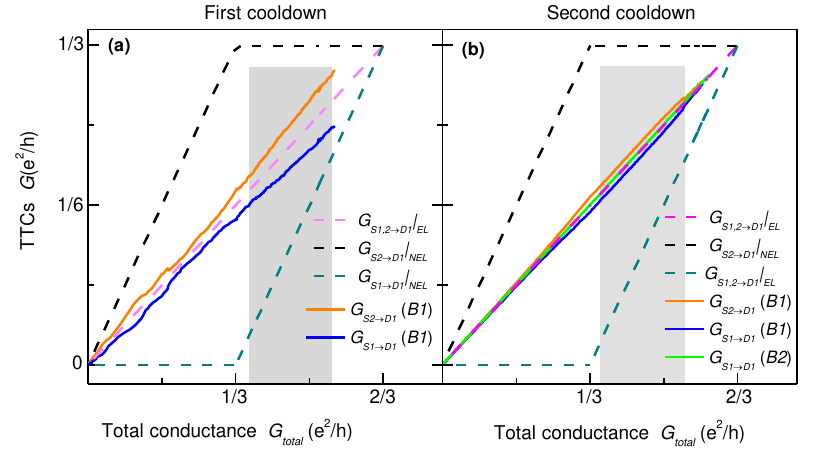}
\centering \caption[ ]
{\label{fig:deviation from equilibration} (color online) (a) Plot of two terminal conductances (TTCs) $G^{~1/3,\nu_2}_{\rm S2 \rightarrow D1}{(2/3, B1)}$ (orange curve) and  $G^{~1/3,\nu_2}_{\rm S1 \rightarrow D1}{(2/3, B1)}$ (blue curve) versus total conductance $G_{total}$ with $B1$ = 13.98 T at first cooldown. The magenta dashed line represents the full equilibration limit (EL) estimated in eqn.\ref{equation:1} and \ref{equation:2}. Deep green and black dashed lines describe the complete non-equilibration limit (NEL) of TTCs $G^{~1/3,\nu_2}_{\rm S2 \rightarrow D1}{(2/3)}$ and $G^{~1/3,\nu_2}_{\rm S1 \rightarrow D1}{(2/3)}$ respectively, formulated in eqn.\ref{equation:3} to \ref{equation:6}. The measured TTCs do not reach EL for $1/3 < \nu_2 < 2/3$ in the gray shaded region. (b) Plot of TTCs vs $G_{total}$ with magnetic fields $B1$ = 13.98 T and $B2$ = 13.4 T for second cooldown for the same sample. The dashed lines represent the EL and NEL values as described above. The orange curve represents measured value of $G^{~1/3,\nu_2}_{\rm S2 \rightarrow D1}{(2/3, B1)}$, blue curve represents  $G^{~1/3,\nu_2}_{\rm S1 \rightarrow D1}{(2/3, B1)}$  with $B1 = 13.98$ T and the green curve represents $G^{~1/3,\nu_2}_{\rm S1 \rightarrow D1}{(2/3, B2)}$ with $B2 = 13.4$ T. The measured TTCs for 13.98 T also do not reach EL for $1/3 < \nu_2 < 2/3$ in the gray shaded region, while the TTC at 13.4 T matches with the EL.}
\end{figure*}

The total conductance $G_{total}$ in the TTC measurement resembles the gate transmission characteristics and it is determined by the filling fraction beneath the gate as $G_{total}\approx \nu_2  e^2/h$. Hence, we can certainly relate the filling fraction $\nu_2$ beneath the gate G2 with the total transmittance conductance. Notably, for incompressible state at $\nu_2$ = 1/3, the above relation is exact. With this knowledge, the measured TTCs $G^{~1/3,\nu_2}_{\rm S2 \rightarrow D1}{\rm (2/3, 13.98 T)}$ and $G^{~1/3,\nu_2}_{\rm S1 \rightarrow D1}{\rm (2/3, 13.98 T)}$ are plotted (pink and blue lines respectively) against the total transmitted conductance $G_{total}$ (equivalent to $\nu_2$) in Figure \ref{fig:deviation from equilibration}(a). 
 
Under full equilibration during co-propagation, the equilibration limit (EL) of the TTCs can be expressed in terms of the product of two consecutive gate transmission probabilities as
\begin{equation} 
G^{~\nu_1,\nu_2}_{\rm S2 \rightarrow D1}{(\nu_b, B)} \mid_{EL} = \frac{(\nu_1 \times \nu_2)}
{\nu_b} \frac{e^2}{h} \ and
\label{equation:1}
\end{equation}
\begin{equation} 
G^{~\nu_1,\nu_2}_{\rm S1 \rightarrow D1}{(\nu_b, B)} \mid_{EL} = \frac{(\nu_b -\nu_1) \times \nu_2}{\nu_b} \frac{ e^2}{h}.
\label{equation:2}
\end{equation}
The EL  for $\nu_1 = 1/3$ of the TTCs are plotted as dashed red line in Figure \ref{fig:deviation from equilibration}(a). 

In case of complete non equilibration of edge modes during co-propagation along the mesa and also under the gate G2, the non-equilibration limit (NEL) of the TTCs can be written for $0 <\nu_2 \le 1/3$ as
\begin{equation} 
G^{~1/3,\nu_2}_{\rm S2 \rightarrow D1}{(2/3,B)} \mid_{NEL} = \nu_2 \frac{ e^2}{h}  \ and 
\label{equation:3}
\end{equation}
\begin{equation} 
G^{~1/3,\nu_2}_{\rm S1 \rightarrow D1}{(2/3,B)} \mid_{NEL} = 0,
\label{equation:4}
\end{equation}
and for the filling fraction range $1/3<\nu_2 \le 2/3$ as,
\begin{equation} 
G^{~1/3,\nu_2}_{\rm S2 \rightarrow D1}{(2/3,B)} \mid_{NEL} =\frac{1}{3} \frac{e^2}{h} \ and 
\label{equation:5}
\end{equation}
\begin{equation} 
G^{~1/3,\nu_2}_{\rm S1 \rightarrow D1}{(2/3,B)} \mid_{NEL} = (\nu_2-1/3) \frac{e^2}{h}.
\label{equation:6}
\end{equation}
The NELs are plotted in black dashed line in Figure \ref{fig:deviation from equilibration}(a) for $G_{\rm S1 \rightarrow D1}$ and deep green dashed line for $G_{\rm S2 \rightarrow D1}$ conductances.

Since the edge modes does not fully equilibrated during co-propagation at high magnetic field, the deviation from the EL line is expected up to filling $\nu_2 = 1/3$ (i.e. $G_{total} = 1/3$) and indeed it is observed in Figure \ref{fig:deviation from equilibration}(a). Interestingly, the measured TTCs for the compressible region in $\nu_2 > 1/3$ (shaded region) do not approach the EL line, instead, they remain separated from each other.

Similarly, the TTCs for the second cooldown are plotted against the total transmitted conductance $G_{total}$ in Figure \ref{fig:deviation from equilibration}(b). Here also the measured TTCs at 13.98 T do not match with the EL line in the shaded region. As the $G_{total}$ value is approaching towards 2/3, the measured TTCs merge together and approach towards the value 1/3. This new result is consistent with our simple model. Notably, the TTCs (green curve) for lower magnetic field (13.4 T) merge with the EL line as expected from full equilibration during copropagation due to very smaller equilibration length compared to propagation length of 125 $\mu$m. Therefore, magnetic field dependent data captures merging of TTCs with EL at low magnetic field and the deviation from EL at higher magnetic fields as well.

Therefore, the result confirms persistence of 1/3 outer mode beneath the gate G2 in the filling fraction range $1/3 <\nu_2 < 2/3$ at higher magnetic fields as shown in the schematic \ref{fig:g1 at 1/3 filling schematics}(a). In absence of edge reconstruction beneath the gate G2 as schematically shown in Figure \ref{fig:g1 at 1/3 filling schematics}(b), both TTCs must reach the EL line in the filling fraction range $1/3 <\nu_2 < 2/3$ for all magnetic field ranges. Hence, our results confirm the existence of a reconstructed 1/3 edge mode in the compressible QH fluid throughout the filling fraction range of $1/3 <\nu_2 < 2/3$.


\section*{Discussion}

The incompressibility of FQH state highly depends on the sample quality. It should be noted that the 2DES mobility decreases with depleting carrier density beneath the gate because of reduced screening of the disorder potential \cite{Rössler2010}. The experimental data shows a clear fractional conductance plateau at 1/3 in gate transmission characteristics as shown in Figure \ref{fig:device diagram and characterization}(c-d). The observation confirms good quality of the 2DES even at reduced density beneath the gate, that helps to carry out the experiments in the filling fraction below 2/3.

Our results confirm that the reconstructed 1/3 downstream edge mode persists for the compressible fluid throughout the filling fraction range of $1/3 <\nu_2 < 2/3$ including the 1/2 filling fraction region which resides in the middle of this filling fraction range. The 1/2 filling state is expected to be a composite Fermions Fermi sea \cite{Halperin1993,Jain2007,Willet1993} in our sample as there is no plateau is observed at the 1/2 filling fraction. Therefore, the composite Fermions Fermi sea at 1/2 filling can host reconstructed 1/3 downstream edge mode as discussed in Figure \ref{fig:schematics of edge reconstruction}(a). Reconstruction of edges surrounding Jain states are modeled by considering the interactions between composite fermions occupying a finite number of Lambda levels each deformed by the edge confinement potential \cite{PhysRevB.84.245104, PhysRevB.86.115127}. The observations here indicates a possible generalization to case of the composite fermion Fermi liquid which involves an infinite number of such composite fermion Lambda levels.

It is interesting to study the edge reconstruction for the incompressible FQH state at 1/2 filling  in special 2DEG with wide quantum well system \cite{Suen1992,Shabani2009,Hasdemir2015,Zhao2021,Sharma2024}. The incompressible FQH state at 1/2 filling arises from pairing of composite Fermions \cite{Zhao2021,Sharma2024,Scarola2002prl} due to short range correlation. Therefore, the edge reconstruction in presence of pairing at 1/2 filling might be different due to competition between long range Coulomb interaction and short range residual correlation. To study the impact of short range residual correlation on edge reconstruction, our experimental methodology is very useful.

In the present 2DEG sample, expected incompressible FQH states of the Jain sequence in between filling fractions 1/3 and 2/3 are not observed, except an unidentified structure above 1/2 conductance. However, throughout the range of filling fraction, 1/3 to 2/3, reconstructed 1/3 fraction edge mode persists. A detailed study of edge equilibration beneath the gate might be performed by separately measuring the currents in the outgoing two modes, which is beyond the scope of this paper.

At magnetic field 14 T highest equilibration length of 104 um at filling 2/3 is obtained. Even at higher magnetic field the system will enter into the compressible region with filling $\nu < 2/3$. Since the compressible fluid possess 1/3 edge mode, it’s equilibration length is expected to be higher at high magnetic field. In such a condition, using the robust 1/3 edge mode multiple concatenated interferometers might be utilized for quantum information processing.

\section*{Conclusions}

In conclusion, we have explained how the fractional edge modes of bulk 2/3 FQH state are connected to the top gate defined fractional compressible fluid in the filling fraction range of 1/3 to 2/3. With this picture, we have demonstrated that a reconstructed fractional $e^2/3h$ edge mode at the boundary of the compressible fluid persists for filling fraction above 1/3. We have also estimated the equilibration length of the fractional mode in our accessible magnetic field range, where equilibration length is increasing with magnetic field.Our studies pave a way to achieve a robust fractional edge mode even in the compressible fluid at higher magnetic fields. Our work will stimulate further study on edge reconstruction of other QH states, including its implications in quantum electron optics.

\section{Acknowledgements}
Authors thank Krishanu Roychowdhury for valuable comments and suggestions.

\bibliography{edgereconstruction_revised}

\end{document}